\documentstyle[aps,epsf,twocolumn]{revtex}
\newcommand{\Ha}{\mbox{$\cal H$}}

\newcommand{\Q}{\mbox{$\cal Q$}}

\begin{document}
\title{ 
Dispersion management in optical fiber links:\\
Integrability in leading nonlinear order}  
\author { Yuri V.  Lvov \cite{yuri}, Ildar R. Gabitov 
\cite{gab}} \address{\cite{yuri}~~
  Center for Nonlinear Studies, Los Alamos National Laboratory, Los
  Alamos, NM 87544\\ 
\cite{gab}~~ Theoretical Division, Los Alamos
  National Laboratory, Los Alamos, NM 87544} 
\maketitle
\begin{abstract}
  { {\bf Abstract} We show that an integro-differential equation model
    for pulse propagation in optical transmission lines with
    dispersion management, is integrable at the {\it leading nonlinear
    order}. This equation can be transformed into the nonlinear
    Schroedinger equation by a near-identity canonical transformation
    for the case of weak dispersion. We also derive  the next order
    (nonintegrable) correction.  }
\end{abstract}

\section{Introduction and Main Ideas} The impressive progress in
high-bit-rate optical fiber communications achieved during the last
few years was primarily due to the invention of the dispersion
management (DM) technique \cite{Lin}, which has become the key
technological component of fiber telecommunications. A recent example
of just how successful this technique is  the
1.02~Tbit/s data transmission achieved in standard optical fiber with
dispersion management over 1000km \cite{Georges}.

The main factor that limits the bit-rate is pulse-broadening due to
the chromatic dispersion of optical fiber. This broadening is
characterized by the dispersion length, $Z_{\rm dis} \sim (d \times
({\rm BR})^{2})^{-1}$. Here, $d$ is the fiber chromatic dispersion and
${\rm BR}$ is the bit-rate. The dispersion length $Z_{\rm dis}$ is the
distance at which the pulse-width approximately doubles due to
dispersive broadening. This distance decreases as a quadratic function
of the bit-rate.
 
The basic idea of the DM technique is to compensate the fiber
chromatic dispersion by periodically incorporating an additional
element, such as a dispersion compensating fiber or fiber chirped
gratings, whose sign of the chromatic dispersion coefficient is 
opposite to that of the main optical fiber. (See Fig.).

\

\

\begin{figure}
\epsfxsize=7cm
\epsffile{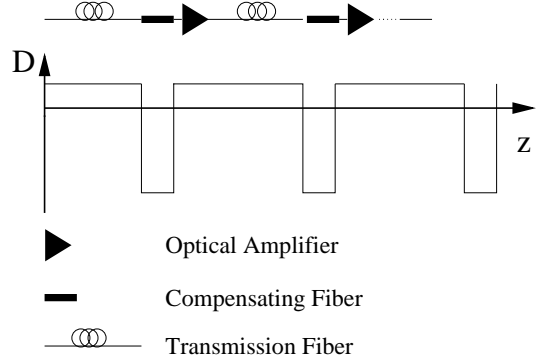}

\

\

\caption{ A schematic of the dispersion management of an optical fiber
link is presented on the upper part of the figure. The dependence of
fiber chromatic dispersion $D$ as a function of the distance $z$ is
presented below.  }
\end{figure}

Modern optical transmission systems must satisfy very strict
requirements for bit-error-rate (BER $\simeq 10^{-12}$ to $10^{-15}$).
Therefore, the pulse amplitude should be large enough so that it can
be effectively detectable. As a consequence, the Kerr nonlinearity of
the fiber refractive index $n=n_{0}+\alpha I$ ($n_{0}$ -- linear part
of the refractive index, $I$ -- pulse intensity, and $\alpha$ --
coefficient of Kerr nonlinearity) should be taken into account. The
spectrum of an optical pulse with characteristic power $P_0$ will
experience noticeable nonlinear distortion at distances grater than
the characteristic nonlinear length $Z_{\rm nl}=(\alpha P_{0})^{-1}$.
A natural idea is to control the nonlinear spectral distortion by
properly choosing the value of the residual dispersion over the period
of the fiber link.

The dispersion management technique is currently  the focus of
intensive experimental, numerical and theoretical research
\cite{Ildar1,Georges,Doran1,Doran2,Doran3,Kawanashi,Menuyk,Ablowitz,Suzuki,Kodama1}
because of  its excellent practical performance. Optical pulses in DM fiber
links exhibit unexpected soliton-like properties, such as high
stability and elastic interaction, despite the fact that the governing
equation does not belong any known class of integrable equations.

In this paper, we demonstrate that the leading-order equation
describing the slow dynamics \cite{Ildar1} of optical pulses in such
systems is close to integrable for the case of weak dispersion management.
We also present the next order (nonintegrable) correction. This correction
is small for { weak} dispersion management.

The problem of weak dispersion management was considered in
\cite{Kodama_hasegawa} for the first time as a theoretical example of
periodic variation of the dispersion coefficient in the model of
transmission link. The power of Lie transform technique proposed by
authors earlier in \cite{Kodama6,Kodama7} was also shown in
\cite{Kodama_hasegawa}. This paper was published long before
dispersion compensation became  popular and before its strength was proven
experimentally. The main result of this paper was that pulse
propagation in  fiber links with weak variation of dispersion can be
described in leading order by an unperturbed nonlinear Schroedinger
equation. Using a formalism which is common in  weak turbulence
theory\cite{ZLF}, we derived similar results and investigated its
validity by calculating the higher order corrections.

\section{Basic Equations}
The relevant equation for the electric field envelope $E(z,t)$ is the
Nonlinear Schroedinger equation (NLS) with periodically varying
dispersion and an external force representing fiber losses and
amplification:
\begin{eqnarray}\label{NLS}
i E_z +\frac{1}{2}\frac {Z_{\rm nl}}{Z_{\rm dis}}d(z) E_{tt} + |E|^2 E = i R'(z) E \\ R'(z)E =
Z_{\rm nl}\left[ -\gamma + r \sum\limits_{k=1}^{\infty} \delta(z-z_k)
\right].
\end{eqnarray}
Here, $\gamma$ describes the fiber losses. The amplifiers that
compensate the fiber losses are placed periodically at points $z_k$
and separated by distances $Z_{\rm a}\sim 1/\gamma$. The coefficient
of amplification $r$ is chosen to exactly compensate energy losses
after passing the amplification distance $Z_{\rm a}$.  The function
$d(z)$ is defined by
\begin{eqnarray} d(z)=  \left[
\begin{matrix}{\ \ 1,\ \  {\rm in\ \ the \ \ 
     transmission\ \  fiber}, \ \ \ \nonumber \\
- d_{comp}/d_{trans}, {\rm in\ \ the \ \ compensating \ \  fiber}}
\end{matrix}.\nonumber \right.
\end{eqnarray}
Without loss of generality, we consider the case when the period of
amplification is equal to the period of compensation. The theory can
be trivially extended to the general case.

In practice, ${Z_{\rm nl}}/{Z_{\rm dis}}\gg 1$ and $\gamma {Z_{\rm nl}}
\gg 1$. Therefore, on the scale of the amplification distance
$Z_{\rm a}$, the pulse dynamics is practically linear. On the other
hand, the linear evolution of the signal preserves the Fourier
spectrum of the pulse. One can consider optical pulse propagation over
one period of a fiber link  to be  a mapping of the input pulse into the
output pulse. To preserve the bit pattern, this mapping must be both
stationary and stable. Stationarity can be achieved by a proper choice
of the amplification coefficient of the amplification and by complete
compensation of dispersion. In this case, stationarity will be
achieved for any pulse shape.  Stability, on the other hand, is more
complicated, and can be investigated through the slow nonlinear pulse
dynamics.

In order to consider slow nonlinear pulse dynamics it is therefore
natural to transform the equations into the Fourier representation,
and study the slow dependence of the spectrum on the coordinate $z$.
One can see the separation of scales in the equation (\ref{NLS}):
$E_z$ is dominated by the RHS and dispersion term. Let us remove the
external force (RHS) from (\ref{NLS}) by the transformation $E =
q(z,t)\exp{({R(z)})}$, and we  obtain the equation
\begin{eqnarray}
i q_z + \frac{1}{2}\frac {Z_{\rm nl}}{Z_{\rm dis}} d(z) q_{tt} + c(z)
|q|^2 q = 0, \end{eqnarray} where $c(z)=\exp{(2{R(z)})}$.  

Let us decompose  $d(z)$ into the sum
\begin{eqnarray}
 d(z) = \langle d(z)\rangle  + \tilde d(z), \ \ 
\langle  \tilde d(z) \rangle =0, \nonumber \\
\end{eqnarray}
where 
$$\langle f(z)\rangle\equiv \frac{1}{Z_{\rm a}}\int\limits_0^{Z_a}
f(z) d z. 
$$
Following \cite{Ildar1}, we represent $q(t,z)$ as
$$
q(z,t) = \int\limits_{-\infty}^{\infty} d \omega a_\omega(z)
\exp{(i \omega t + i \omega^2 \frac{1}{2}\frac {Z_{\rm nl}}{Z_{\rm
      dis}}\int_0^z \tilde d(z) d z )},$$
and compute the equation for
the slow dynamics of the pulse's Fourier spectrum in Hamiltonian
form
\begin{equation} 
\frac{\partial a_{\omega}}{\partial z} + i\frac{\delta \Ha}{\delta 
a_{\omega}^*}=0.
\label{CanonicalEquation}
\end{equation}
with the Hamiltonian
\begin{eqnarray} \label{Hamilton} 
{\Ha}= \int \frac{1}{2}\frac{Z_{\rm nl}}{Z_{\rm dis}} 
 <d(z)> \omega^2 |a_{\omega}|^2 d \omega+ 
\nonumber
\int d
\omega_1 d \omega_2 d \omega_3 d \omega_4 
\nonumber\\ \times
F^{ \omega_1
\omega_2}_{\omega_3 \omega_4} \delta{( \omega_1 +\omega_2 -\omega_3
-\omega_4 )} a_{\omega_1}^*  a_{\omega_2}^* a_{\omega_3 } a_{\omega_4},
\label{Orig}
\end{eqnarray}
where 
$$
F^{ \omega_1 \omega_2}_{\omega_3 \omega_4} = c(z)\exp{(i (\omega_1^2
  +\omega_2^2 -\omega_3^2 -\omega_4^2 ) \frac{1}{2}\frac {Z_{\rm
  nl}}{Z_{\rm dis}} \int_o^z\tilde d(\xi) d \xi)}.$$ Note that the
  kernel $F^{ \omega_1 \omega_2}_{\omega_3 \omega_4}$ depends only on
  the composed variable $g=(\omega_1^2 +\omega_2^2 -\omega_3^2
  -\omega_4^2 )$, which,  as we will see later, leads to important
  consequences.

The leading-order approximation for the slow nonlinear dynamics can be
obtained by averaging (\ref{Hamilton}) over $Z_a$. The resulting
equation is valid if the pulse spectrum does not change appreciably
over the amplification distance. The next order correction to the
averaged equation was calculated in \cite{IldatTurBurtsev} by using a
Lie transform \cite{Kodama6,Kodama7}.  An alternative method for calculating
these corrections was proposed in \cite{ZakharovP}.

In the simple case of communication through one frequency channel,
with the additional assumption that pulses do not interact with each
other, one can use the quasi self-similar character of the
single-pulse behavior \cite{Kumar}, which was proposed for the first
time by Talanov (lens transformation)\cite{Talanov}. As a result, the
slow dynamics of a single pulse is described by the Nonlinear
Schroedinger equation with a quadratic potential. This approach was
extensively discussed in \cite{Kodama1},
\cite{Kumar},\cite{GabitovTuritsyn}, and allows one to
explicitly take into account the evolution of the single pulse spectra
within the system period. Another advantage of the quasi
self-similarity is that it also holds for very small values of the
residual dispersion.  It was shown in \cite{Kodama1},\cite{Doran} that
stable single pulse propagation is possible for zero or even negative
residual dispersion for special configurations of the dispersion map.

However Talanov's method cannot be applied to the case of several
interacting pulses within one frequency channel, nor can it describe
the evolution of an optical pulse propagating in a system that
utilizes the wavelength division multiplexing (WDM) technique. The WDM
technique exploits several frequency channels for data transmission.
Optical pulse streams in systems that use this technique propagate in
every channel with their own velocities and interact through the
common refractive index.

In the fiber links with dispersion management, the nonlinear effects
are smaller than the dispersive effects, therefore the Fourier
spectrum of a pulse does not change much between amplifiers. Hence,
equation (\ref{CanonicalEquation}) can be replaced by the equation
averaged between the amplifiers. In the corresponding averaged
Hamiltonian, $F^{ \omega_1 \omega_2}_{\omega_3 \omega_4}$ is replaced
by $ T^{ \omega_1 \omega_2}_{\omega_3 \omega_4} = <F^{ \omega_1
\omega_2}_{\omega_3 \omega_4}>$, so that this Hamiltonian becomes
\begin{eqnarray} \label{HamiltonTran} 
{\Ha}&=& \int \frac{1}{2}\frac{Z_{\rm nl}}{Z_{\rm dis}} <d(z)>
\omega^2 |a_{\omega}|^2 d \omega+ \nonumber \int d \omega_1 d \omega_2
d \omega_3 d \omega_4 \nonumber\\ &&\times T^{ \omega_1
\omega_2}_{\omega_3 \omega_4} \delta{( \omega_1 +\omega_2 -\omega_3
-\omega_4 )} a_{\omega_1}^* a_{\omega_2}^* a_{\omega_3 } a_{\omega_4},
\nonumber \\ {\rm here} \nonumber \\ T^{ \omega_1 \omega_2}_{\omega_3
\omega_4} &\equiv& \frac{1}{z_a}\int_0^{z_a} d z c(z) \nonumber\\
&&\times\exp{{(i (\omega_1^2 +\omega_2^2 -\omega_3^2 -\omega_4^2
)\frac{1}{2}\frac {Z_{\rm nl}}{Z_{\rm dis}} \int_o^z\tilde d(\xi) d
\xi)}},
\end{eqnarray} 
This averaging is equivalent to assuming that higher order corrections
in the Lie expansion are small compared to the leading order.

\section{Integrability at the leading order}
In our further analysis we will utilize ideas that  are well
developed in the theory of weak turbulence \cite{ZLF}.  We
rewrite the Hamiltonian (\ref{HamiltonTran}) as
\begin{eqnarray} 
\Ha = \int \Q_{\omega} a_{\omega} a_{\omega}^* d{\omega} + \nonumber\\
\int T^{{\omega}_1 {\omega}_2}_{{\omega}_3 {\omega}_4}
a_{{\omega}_1}^*a_{{\omega}_2}^*a_{{\omega}_3}a_{{\omega}_4}
\delta_{{\omega}_1+{\omega}_2-{\omega}_3-{\omega}_4}d{\omega}_1
d{\omega}_2d{\omega}_3d{\omega}_4,\cr
\label{Hamiltonian}
\end{eqnarray}
where we define the leading order quadratic dispersion relation as
\begin{eqnarray} 
\Q_{\omega}=\beta <d(z)> {\omega}^2, \ \ \ 
\beta\equiv \frac{1}{2}\frac {Z_{\rm nl}}{Z_{\rm dis}}, 
\label{dispersion} \end{eqnarray}
\noindent Then  $a_{\omega}$  satisfies the canonical  equation of motion:
\begin{equation} 
\frac{\partial a_{\omega}}{\partial z} + i\frac{\delta \Ha}{\delta a_{\omega}^*}=0.
\label{CanonicalEquation2}
\end{equation}
If $T^{{\omega}_1 {\omega}_2}_{{\omega}_3 {\omega}_4}$ is a constant
independent of the $\omega$'s, then equations  
(\ref{Hamiltonian}-\ref{CanonicalEquation2}) correspond to the
usual NLS equation.  With $T^{{\omega}_1 {\omega}_2}_{{\omega}_3
{\omega}_4}$ given by (\ref{HamiltonTran}) \cite{Ildar1}, these 
equations  describe the averaged, slow-time evolution of an optical
pulse propagating in a fiber-optical link with periodic amplifiers,
damping, and dispersion compensators.
 
To find the leading-order approximation to
(\ref{Hamiltonian}-\ref{CanonicalEquation2}),
we follow \cite{ZLF,Z74,KRS90,KRS94} and perform the transformation

\begin{eqnarray} \label{transformation}
a_{\omega} &=& b_{\omega} + \int B^{{\omega} {\omega}_1}_{{\omega}_2
{\omega}_3}b^*_{{\omega}_1}b_{{\omega}_2}b_{{\omega}_3}\delta_{{\omega}+
{\omega}_1-{\omega}_2-{\omega}_3}+\nonumber\\&& {\rm{\ higher \ \
order\ \ interactions}}.
\end{eqnarray}
The transformation (\ref{transformation}) is canonical up to
and including the cubic terms if the coefficients 
$B^{{\omega} {\omega}_1}_{{\omega}_2 {\omega}_3}$ 
satisfy the symmetry condition \cite{ZLF}
\begin{eqnarray} \nonumber
B^{{\omega} {\omega}_1}_{{\omega}_2 {\omega}_3} = 
B^{{\omega}_1 {\omega}}_{{\omega}_2 {\omega}_3} =
B^{{\omega} {\omega}_1}_{{\omega}_3 {\omega}_2} = 
-( B^{{\omega}_2 {\omega}_3}_{{\omega} {\omega}_1})^*.
\end{eqnarray}
Substitution of (\ref{transformation}) into the Hamiltonian (\ref{Hamiltonian}) 
transforms (\ref{Hamiltonian}) into 
\begin{eqnarray}
\Ha &=& \int \Q_{\omega} b_{\omega} b_{\omega}^* d{\omega} + \cr&&
\int \left(T^{{\omega}_1 {\omega}_2}_{{\omega}_3
{\omega}_4} +
(\Q_{\omega}+\Q_{{\omega}_1}-\Q_{{\omega}_2}-\Q_{{\omega}_3})
B^{{\omega}_1 {\omega}_2}_{{\omega}_3 {\omega}_4}\right) \nonumber\\ &&
b_{{\omega}_1}^*b_{{\omega}_2}^*b_{{\omega}_3}b_{{\omega}_4}
\delta_{{\omega}_1+{\omega}_2-{\omega}_3-{\omega}_4}d{\omega}_1d{\omega}_2d{\omega}_3d{\omega}_4\cr
&+&  {\rm{\ higher \ \ order\ \ interactions}}.
\label{HamiltonianTransformed}
\end{eqnarray}

The main mechanism of wave interaction
is four-wave scattering, which occurs when $\omega$'s and
$\Q_\omega$'s satisfy the following resonant conditions
\begin{eqnarray} \label{res4}
\Q_{{\omega}} + \Q_{{\omega}_1} &=& \Q_{{\omega}_2} + \Q_{{\omega}_3}
\cr {\omega} + {\omega}_1 &=& {\omega}_2 + {\omega}_3.
\label{ResonanceManifold}
\end{eqnarray}
The set of $\omega$'s that satisfies these conditions is called the
{ resonant manifold}.  The form of the transformed Hamiltonian 
(\ref{HamiltonianTransformed})
explains the importance of the notion of a resonant manifold.  Indeed,
at the points where (\ref{ResonanceManifold}) holds, in general, one
cannot choose the arbitrary coefficients $B^{{\omega}_1
{\omega}_2}_{{\omega}_3 {\omega}_4}$ to exclude the terms
$T^{ \omega_1  \omega_2}_{\omega_3 \omega_4}
a_{\omega_1}^* a_{\omega_2}^* a_{\omega_3} a_{\omega_4}$ from the
Hamiltonian because of ``small resonant denominators''.

For the case of quadratic dispersion (\ref{dispersion}), equations
(\ref{ResonanceManifold}) have only { trivial} solutions
\begin{eqnarray} \label{Triv4}
{\omega}_2 &=& {\omega}_1,\hspace{.5cm} {\omega}_3 = {\omega},\cr
or\hspace{.5cm}{\omega}_3 &=& {\omega}_1,\hspace{.5cm} {\omega}_2={\omega}.
\end{eqnarray}

In this case the value of $T^{{\omega}
{\omega}_1}_{{\omega_3}{\omega}_4}$ on the resonant manifold
(\ref{ResonanceManifold}),(\ref{Triv4}) is equal to a constant
independent of $\omega$'s: $T^{{\omega} {\omega}_1}_{{\omega}
{\omega}_1} =T_0$.  We will show below that the leading nonlinear
approximation the equation (\ref{CanonicalEquation2}) with
Hamiltonian (\ref{Hamiltonian}) and quadratic dispersion relation
(\ref{dispersion}) is the NLS, and hence it is { integrable}. Thus, we
can { integrate} (\ref{CanonicalEquation2}) at the { leading nonlinear
order}.

For the case of interest, the interaction matrix element is a
function of $g\equiv (\omega_1^2 +\omega_2^2 -\omega_3^2
-\omega_4^2)$:
$$T^{ \omega_1 \omega_2}_{\omega_3 \omega_4} = T(g).$$ Now we choose the
canonical transformation kernel as
\begin{eqnarray}
B^{ \omega_1 \omega_2}_{\omega_3 \omega_4}=-\frac{ T^{ \omega_1
\omega_2}_{\omega_3 \omega_4} - T^{\omega_1 \omega_1}_{\omega_1
\omega_1}} { \omega_1^2 +\omega_2^2 -\omega_3^2 -\omega_4^2},
\end{eqnarray}
or equivalently,
\begin{eqnarray}
B^{ \omega_1 \omega_2}_{\omega_3 \omega_4}=-\frac{T(g) -T(0)}{g}.
\label{TransformationKernel}
\end{eqnarray}
Note that this expression does not have a ``small denominator'' problem.
It is not singular on trivial resonances, i.e. it is not singular
for $g=0$.  This fact is obvious from an expansion of $T^{ \omega_1
\omega_2}_{\omega_3 \omega_4}$ in powers of $g$:
\begin{eqnarray}
T(g)=T(g=0) + g \frac{\partial T}{\partial g} +
\sum\limits_{k=2}^{\infty}\frac{g^k}{k!}  \frac{\partial T^{ \omega_1
\omega_1}_{\omega_1 \omega_1}}{\partial g}.
\label{Taylor}
\end{eqnarray}
Here $T^{\omega_1 \omega_1}_{\omega_1 \omega_1} \equiv T(g=0)\equiv
T_0$, i.e.  a constant independent of $\omega_i$.  The transformed
Hamiltonian acquires the form
\begin{eqnarray}
\label{TransformedToNLS}
\Ha = \int \Q_{\omega} b_{\omega} b_{\omega}^* d{\omega} + \cr \int
 T_0 b_{{\omega}}^*b_{{\omega}_1}^*b_{{\omega}_2}b_{{\omega}_3}
 \delta_{{\omega}+{\omega}_1-{\omega}_2-{\omega}_3}d{\omega} d
 {\omega}_1 d {\omega}_2 d {\omega}_3 \cr + {\rm{\ higher \ \ order\ \
 interactions}},
\end{eqnarray}
which is the Hamiltonian of the NLS equation to leading order.  Thus
equation (\ref{CanonicalEquation2}) with Hamiltonian
(\ref{Hamilton}) and quadratic dispersion relation (\ref{dispersion})
is integrable at the leading nonlinear order.


The condition of quasi-identity of transformation
(\ref{transformation}) reads as
\begin{eqnarray} 
B^{ \omega_1 \omega_2}_{\omega_3
\omega_4} \ll 1. \label{Condition}
\end{eqnarray}
Let us estimate the $B^{ \omega_1 \omega_2}_{\omega_3 \omega_4}$ in
(\ref{TransformationKernel}).  For the case of a lossless piece-wise
constant dispersion map with alternating pieces of fibers of length
$l_1$ with dispersion $<d> + d_1$ and of length  $l_2$ with
dispersion $<d> + d_2$, ($l_1 d_1 + l_2 d_2 = 0$), $T^{ \omega_1
\omega_2}_{\omega_3 \omega_4}$ acquires the form
\begin{eqnarray} 
T^{ \omega_1 \omega_2}_{\omega_3 \omega_4} &=& 
\frac{\sin (\delta g )}{\delta g}, \label{SINUS}
\end{eqnarray}
where $\delta$ is the  strength of a dispersion map. 
Thus condition (\ref{Condition}) becomes
\begin{eqnarray} 
\delta^2 \ll  \langle d \rangle, 
\label{Condition2}
\end{eqnarray}
so that the transformation (\ref{transformation}) is quasi-identical
for the so called { weak} dispersion maps, i.e. for the cases of
small variation of the dispersion on top of big dispersion.


In this paper, we only study the sixth-order interaction in the
averaged equations.  V.E. Zakharov \cite{ZakharovP} recently pointed out
that a canonical transformation similar to (\ref{transformation}) can also 
be used to bring the original nonaveraged problem (\ref{Orig}) into a form
analogous to that developed here for the averaged equation, namely,
the Nonlinear Schroedinger Equation with higher-order corrections.  We
believe that the present work is an important stepping stone in the
proper development of a  more sophisticated transformation, which  will
give us sufficient understanding to  attack this much harder
problem.

\section{Six-wave interactions on a resonant surface}
In this section, we calculate the  sixth-order interaction matrix
element for the Hamiltonian (\ref{TransformedToNLS}). The Hamiltonian
(\ref{TransformedToNLS}) is the averaged Hamiltonian, and, including
sixth-order terms, has the form:
\begin{eqnarray} 
\Ha &=& \int d  \omega\Q_\omega |b_\omega|^2 + \int
T_0 d \omega d \omega_1 d \omega_3 d \omega_4 b_{\omega_1}^* b_{\omega_2}^* 
b_{\omega_3}b_{\omega_4} \cr &+& \int d \omega d \omega_1
d \omega_2 d \omega_3 d \omega_4 d \omega_5 
T^{\omega,\omega_1,\omega_2}_{\omega_3,\omega_4,\omega_5} 
\nonumber\\ &&
b_{\omega}^*b_{\omega_1}^*
b_{\omega_2}^* b_{\omega_3} b_{\omega_4} b_{\omega_5} \delta ^{\omega+\omega_1+\omega_2}_{\omega_3+\omega_4+\omega_5},\cr
&&\label{BirghoffHIGH}
\end{eqnarray}

There are two ways of calculating the sixth-order interaction matrix
element $T^{\omega,\omega_1,\omega_2}_{\omega_3,\omega_4,\omega_5} $
for the Hamiltonian (\ref{BirghoffHIGH}). The first way is to
calculate the higher order terms in the canonical transformation
(\ref{transformation}). The second way, which is simpler and clearer,
is to compute $T^{\omega \omega_1 \omega_2}_{\omega_3 \omega_4
\omega_5}$ involving standard Wyld diagram formalism \cite{Wyld} which
is a generalization of the Feynman diagram technique for classical
nonequilibrium systems.

Following \cite{FIVE} the matrix element $T^{\omega \omega_1
  \omega_2}_{\omega_3 \omega_4 \omega_5}$ can be represented as sum
of all topologically different irreduceable diagrams with no internal
loops. Below we present, an example of such a diagram. The structure
of this diagram can be understood as follows: This diagram consists
of two vortices A and B. Each of these vortices forms a  diagram
which corresponds to four wave mixing. For computing of six order
interaction term in the perturbation theory one should take two
four-wave vortesces, connect them using all topologically possible
configurations and then summ all of them. The number of such
nonequvalent configurations is nine. The diagram presented below is
one one of these nine terms.

\begin{figure}
\epsfxsize=7cm
\epsffile{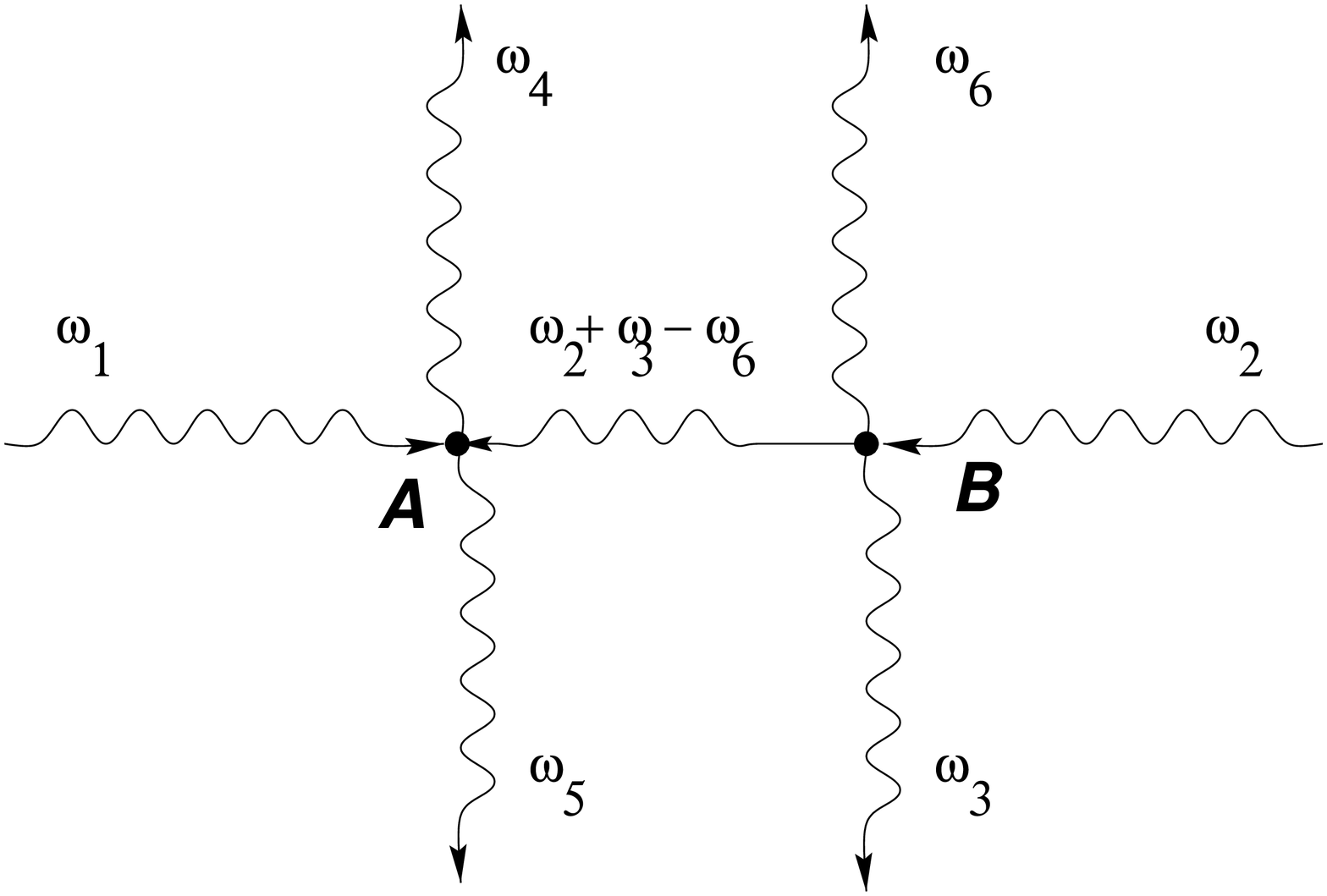}

\end{figure}

\

\

\

\

\

\

The  analytical expression corresponding to this diagram is
$$
\frac{ T^{\omega_1,\omega_2+\omega_3-\omega_6}_{\omega_4 \omega_5} T^{\omega_2 \omega_3}_{\omega_6, \omega_2+\omega_3 -\omega_6 }}
{\Q_{\omega_2}+\Q_{\omega_3}-\Q_{\omega_6}-\Q_{\omega_2+\omega_3-\omega_6}}.
$$

After summing up all 9 diagrams, the sixth order matrix element $T^{\omega
  \omega_1 \omega_2}_{\omega_3 \omega_4 \omega_5}$ can be represented
as
\begin{equation}
T^{{\omega_1} {\omega_2} {\omega_3} }_{{\omega_4} {\omega_5}
{\omega_6}}=
\sum\limits_{i=1}^{i=9}\frac{|T(g_i)|^2}{g_i},
\end{equation} 
where the nine variables $g_i,\ \ \ i=1,..,9$ are given  by the
following expressions \\
\begin{eqnarray}
g_i=\{
   -2({\omega_4}-{\omega_1})({\omega_4}-{\omega_2})  ,
  -2({\omega_4}-{\omega_1})({\omega_4}-{\omega_3}), \nonumber\\ \nonumber 
  -2({\omega_4}-{\omega_2})({\omega_4}-{\omega_3}),
  -2({\omega_5}-{\omega_1})({\omega_5}-{\omega_2}),  \nonumber\\ \nonumber
  -2({\omega_5}-{\omega_1})({\omega_5}-{\omega_3}),
  -2({\omega_5}-{\omega_2})({\omega_5}-{\omega_3}),  \nonumber\\ \nonumber
  -2({\omega_6}-{\omega_1})({\omega_6}-{\omega_2}),
  -2({\omega_6}-{\omega_1})({\omega_6}-{\omega_3}),\nonumber\\ \nonumber 
  -2({\omega_6}-{\omega_2})({\omega_6}-{\omega_3}) \}.  \end{eqnarray}
  
The resulting expression can be viewed as the measure of the
``distance'' of the perturbed  system from its integrable analog.
Indeed direct calculations show that for the case of NLS, when
$T^{\omega_1,\omega_2}_{\omega_3,\omega_4}=const$, this sixth order
matrix element is identically equal to zero on the resonant manifold
\begin{eqnarray} \label{res6}
\Q_{\omega} + \Q_{\omega_1} +\Q_{\omega_2}&=& \Q_{\omega_3} + \Q_{\omega_4}
+\Q_{\omega_5}\cr \omega +  \omega_1 + \omega_2 &=& \omega_3 + 
\omega_4+ \omega_5.
\end{eqnarray}
In order to bring equation (\ref{Hamiltonian}) as close
to the integrable limit as possible, we must minimize $T^{\omega \omega_1
  \omega_2}_{\omega_3 \omega_4 \omega_5}$ over all the tunable system
parameters.

For the case of a lossless piece-wise constant dispersion map, when the fourth
order matrix element is given  by (\ref{SINUS}), we find
(using  { Mathematica 3.0}), that the six wave interaction
matrix element
$$T^{\omega \omega_1 \omega_2}_{\omega_3 \omega_4
  \omega_5} \propto \delta^2 / \langle d \rangle, 
$$
on the 6-wave resonant manifold (\ref{res6}). Therefore the above
perturbation technique is valid for the case of a weak dispersion
map. In leading order, the slow pulse evolution in optical
fiber is governed by the integrable NLS equation.

\section{ Conclusion} 
We showed that the leading-order equation describing pulse dynamics in
optical fiber links with a weak dispersion map is close to integrable.
Using canonical transformations, we found a nonintegrable higher order
correction to the  integrable part of the model equation describing
the slow evolution of optical pulses.

\section{Acknowledgment} We are very grateful to Roberto Camassa,
Gary Doolen, Gregor Kovacic, Victor Lvov and Vladimir Zakharov for
helpful discussions. This work was supported by DOE contract
W-7-405-ENG-36 and the DOE Program Applied Mathematical Sciences
KJ-01-01.

 \end{document}